\documentstyle[12pt,epsf]{article}
\title{On Charged Mesoscopic Metallic Bubbles
    \footnote{This work is partly  supported by the Polish Committee of
          Scientific Research under contract No.  2P3 03B 49 09}}

\author{by \\
              Krzysztof Pomorski \\
   Theoretical Physics Department, University M.C.S., Lublin, Poland \\
                             and \\
               Klaus Dietrich \\
 Physik Department of the Technische Universit\"at M\"unchen, \\
Garching (Fed. Rep. Germany)}

\date{}

\begin{document}
\maketitle

\begin{abstract}
We investigate the existence of stable charged metallic bubbles
using the shell correction method. We find that for a~given
mesoscopic system of $n$ atoms of a~given metal and $q \ll n$
(positive) elementary charges, a~metallic bubble turns out to
have a~lower total energy than a~compact spherical cluster,
whenever the charge number $q$ is larger than a~critical charge 
number $q_c$.
For a magic number ($n-q$) of free electrons, the spherical
metallic bubble may become stable against fission.\\
{\bf PACS:} 36.40.+d, 36.20.Kd, 31.10.+z, 21.90.+f, 21.60.Cs
\end{abstract}

\section{Introduction}

Neutral and charged metal clusters consisting of a~few 100 to
a~few 1000 atoms possibly containing a~limited member
($\leq$~10) of positive or negative surplus charges received
considerable scientific interest since 1984. At that time, when
studying the formation of tiny alkali clusters out of metal
vapour, an enhanced production of clusters with certain atomic
numbers ($n$ = 8, 20, 40, 58, 92, etc.) was observed [1, 2] and
was correlated with the appearance of shell closures for the
motion of the free electrons i.e. the conduction (valence)
electrons in a~spherically symmetric average potential for the
itinerant electrons. A~free electron feels an average potential
which is produced by the background of positive ions, on the one
hand, and by the other free electrons, on the other.

The most prominent peaks in the mass yield were shown to be due
to shell closures in a~spherically symmetric potential [3--8],
whereas tinyer details of the abundance curve were successfully
related to secondary shell effects in axially symmetric
deformed [9--11] and non-axially symmetric deformed [12]
potentials. 

The calculations are usually performed using the Strutinsky
shell correction method [13] or the more involved
self-consistent field approach [14, 15].

A~large amount of beautiful experimental work [16, 17] has been
performed since the discovery in 1984. The comparison between
the experimental and theoretical work (see for instance Ref. 10)
is in general satisfactory. The greatest part of the work has
hitherto been devoted to uncharged clusters. Experiments on
charged metallic clusters were performed by C. Br\'echignac et
al. Li$^{+q}_n$ ($n$ = nr of atoms, $q$ = nr of elementary
charge units) in Ref. 18 and for Sb$^{+q}_n$ in Ref. 19.
It was found that in most cases singly charged clusters are
stable, i.e. their dissociation is endothermic. Clusters with
a~positive charge $q > 1$ are observed if the number $n$ of
constituant atoms is larger than a~critical number $n^q_b$ 
which depends on the system considered. Decay by fission and by
evaporation of neutral or charged fragments compete with each
other, the decay by fission being delayed with respect to the
decay by evaporation.

Theoretical studies of the decay were published by F. Garcias et
al. [20] using a~semi-empirical model for the fission of
multiply charged metal clusters and by D.~Gross [21].

In all the theoretical studies of mesoscopic metallic clusters
it was assumed that the groundstate of the cluster corresponds
either to a~compact spherical or to a compact deformed shape.

As we show in this paper, for large enough charge $q$, the state
of lowest energy of a~charged metallic cluster may correspond to
a~spherical bubble. Within the liquid drop approximation, the
spherical bubble solutions turn out to be unstable versus
fission. This has been shown in nuclear physics quite some time
ago [22]. In a~recent work [23], we found that spherical nuclear
bubbles may be stabilized by shell effects. In this paper, we
show that the same is true for charged metal clusters if the
number of valence electrons corresponds to a~closed shell.

In Section 2, we define the theoretical model and in Section 3
we present the results we obtained. In Section 4 contains
a~short summary and a~discussion of open questions.

\section{Theoretical model}

For mesoscopic metallic clusters containing from 100 
to more than 1000 atoms, the distribution 
of the positively charged ions can be approximately
described by a~homogeneously smeared-out density
which in the simplest case is given by a~step
function. For a~bubble with inner radius $R_2$ and outer 
radius $R_1$, the density of positive ions is thus given by
$$
\rho_{\rm ion}(r) = \,\hskip -3pt\buildrel\circ\over\rho_{\rm ion}
\theta_0(r - R_2)  \, \theta_0(R_1 - r) \,\, ,
\eqno[2.1]
$$
$$
\theta_0(x) = \left\{
\begin{array}{lll}
 1 & {\rm for} & x > 0 \\
 0 & {\rm for} & x < 0
\end{array}\right. \,\, .
$$
The constant bulk density $ \hskip
-3pt\buildrel\circ\over\rho_{\rm ion} > 0$ is usually given in
terms of the radius $\hskip -3pt\buildrel\circ\over r_s$ of a~sphere 
which contains 1 atom on the average
$$
\hskip -3pt\buildrel\circ\over r_s\, = \, [3/4\pi 
     \hskip  -3pt\buildrel\circ\over\rho_{\rm ion}]^{1/3} \,\, .
\eqno(2.2)
$$
For a given number $n$ of atoms, the volume of bubble layer has to
be equal to the one of a~compact spherical cluster of radius
$R_0$ 
$$
{4\pi\over 3}(R^3_1 - R^3_2)\hskip-3pt\buildrel\circ\over\rho_{\rm ion}
= {4\pi\over 3} R^3_0 \, \hskip-3pt\buildrel\circ\over\rho_{\rm ion}
= n \,\, .
\eqno(2.3)
$$
Consequently, the shape of the spherical bubble is determined by
only one free parameter. We choose it to be the ratio
$$
f: = R^3_2/R^3_1
\eqno(2.4)
$$
between the volume of the inner hole to the volume of the entire
bubble.

The conduction electrons move independently in an average
potential $V(r)$ which represents the mean interaction of
a~given electron with all the other electrons and with the
positive back ground charge $\rho_{\rm ion}(r)$. In the
Hartree-Fock approximation the single particle states
$\varphi_\nu(\vec r)$ of the electrons\footnote{We leave away an
explicit notation of the spin degrees because spin-dependent
interactions are neglected.} and the corresponding single particle
energies $\varepsilon_\nu$ are obtained as the selfconsistent
solution of the coupled equations
$$
\left[-{\hbar^2\over 2m}\triangle + \widehat V(r)
\right] \cdot \varphi_\nu(\vec r) = \varepsilon_\nu\varphi_\nu(\vec r) \,\, , 
\eqno(2.5)
$$
$$
 \widehat V(r) \cdot \varphi_\nu(\vec r) = [V_{\rm ion} (r) -
   \widehat V_e(r)] \cdot \varphi_\nu(\vec r) \,\, ,
\eqno(2.6)
$$
$$ 
V_{\rm ion}(r) = -\int dr' {\rho_{\rm ion}(r')\over |\vec r -
   \vec r'|} \,\, ,
\eqno(2.7) 
$$
$$
\widehat V_e(r) \cdot \varphi_\nu(\vec r) = \sum_{\kappa\neq\nu}
\,  n_\kappa \, \int d^3{r'} \, \varphi^{\dagger}_\kappa(\vec {r'}) 
 {e^2_0\over |\vec r - \vec {r'}|} \left[\varphi_\kappa(\vec {r'})
\varphi_\nu(\vec r) - \varphi_\nu(\vec {r'})\varphi_\kappa(\vec r)\right] 
\,\, .
\eqno(2.8)
$$
In (2.8), $e_0$ is the elementary charge and $n_\kappa$ is the
occupation number of the single particle state $\varphi_\kappa$.
For temperature $T=0$, $n_\kappa$ is given by 
$$
n_\kappa = 1~~~~{\rm for}~~~~\varepsilon_\kappa <\varepsilon_F \,;~~~~~~~~~~ 
n_\kappa = 0~~~~{\rm for}~~~~\varepsilon_\kappa > \varepsilon_F \,\, ,
\eqno(2.9)
$$
where $\varepsilon_F$ is the Fermi energy. We assume that the
temperature is zero.

The Hartree-Fock potential (2.8) is seen to be state-dependent
mainly due to the exchange term. Usually, the exchange term is
replaced by a~local density approximation [3, 24]. If we neglect
the exchange term altogether and suppress the exclusion of the
state $\varphi_\kappa = \varphi_\nu$ in the remaining Hartree
potential, we obtain
$$
 \widehat V_e(r)\varphi_\nu(\vec r) = \int d^3 {r'}
{\rho^H_e({r'})e_0\over |\vec r - \vec {r'}|} \,\, ,
\eqno(2.10)
$$
$$
\rho^H_e({r'}) = \sum_\kappa n_\kappa \, e_0 \varphi^\dagger (\vec
{r'})\, \varphi_\kappa (\vec {r'}) \,\, .
\eqno(2.11)
$$
The solution of the remaining set of selfconsistent Hartree
equations (Eqs. (2.5) with (2.6), (2.7), (2.10), (2.11)) is
still a~considerable technical problem. It has been carried through
for instance in Ref. [3].

A considerable simplification is obtained if the selfconsistent
average potential (2.6) is replaced by a phenomenological
ansatz. Most of the shell structure calculations for compact
metal clusters have been performed on the basis of simple
phenomenological potentials, in particular the Nilsson potential
[9], the Saxon-Woods potential [7, 11], and the ,,wine-bottle''
potential [7]. It can indeed be seen from the results of Ref.
[3] that for atom numbers $n \geq 40$ the selfonsistently
calculated potential $V(r)$ resembles a~Saxon-Woods potential
and roughly even to a~square well. We, therefore, felt justified
to represent the average potential $V(r)$ for the case of
a~bubble cluster by an infinite square well with the boundaries
given by the distribution of the positive ions
$$
    V(r) = \left\{
\begin{array}{ll}
 -V_0 \, & {\rm for} R_2 < r < R_1 \\
 +\infty & {\rm otherwise}
\end{array}\right. \,\, .
\eqno(2.12)
$$
This simple choice of the shell model potential has the great
advantage that the eigenfunctions of the Schr\"odinger equation
(2.5) are linear combinations of spherical Bessel- and Neumann
functions and the eigenenergies $\varepsilon_\nu$ are easily
obtained from the boundary conditions at $r = R_{1,2}$ [23].

The well depth $V_0$ in (2.12) is of the order of 0.5 Ry (1 Ry =
13,6~eV). Its value is not relevant for our results because the
shell correction energy (see Eq. 2.16) turns out to be
independent of the constant $V_0$.

We now have to determine the total energy $E$ of the metal
cluster as a~function of the variable $f$ (see Eq. (2.4)). Given
the fact that the ion density $\rho_{\rm ion}(r)$ can be
considered to be constant inside the matter distribution (see
(2.1)), we may write the energy of the system as a~sum of the
energy $E_{\rm LD}$ of a ,,liquid drop'' and a~,,shell
correction energy'' $E_{\rm shell}$ following Strutinsky [13] 
$$ 
E_{\rm tot} = E_{\rm LD} + E_{\rm shell} 
\eqno(2.13) 
$$ 
The liquid drop energy
can be written as a sum of a (negative) term proportional to the
volume ${\rm V}$ of the system and a~(positive) term
proportional to the surface $S$.

Since we consider (positively) charged clusters, we have to add
the electrostatic energy $E_{\rm Cb}$ of the sytem 
$$ 
E_{\rm LD} = -\tau{\rm V} + \sigma S + E_{\rm Cb} 
\eqno(2.14)
$$ 
which turns out to be a much bigger term than in a neutral cluster.

Some authors [10] include a term proportional to the average
curvature of the surface which is the 3$^{\rm rd}$ term in the
expansion of the energy-density functional of a~leptodermons
sytem in terms of $n^{-1/3}$. In the case of a~bubble shape, the
curvature terms arising from the inner and outer surface have
opposite signs beside the fact that their absolute value is much
smaller than the corresponding surface term. We omit the
curvature term thereby following Ref. [21]. Consideration of 
the curvature term would favour spherical bubbles as compared to
compact spheres.

The ,,macroscopic'' electrostatic energy $E_{\rm Cb}$ is given
as a~function of smooth density distributions of the positive
and negative charge by 
$$
    E_{\rm Cb} = {1\over 2} \int d^3r \int d^3{r'} 
   {[\rho_e(r) - \rho_{\rm ion}(r)][\rho_e({r'}) - \rho_{\rm
     ion}({r'})] \over |\vec r - \vec {r'}|}
\eqno(2.15)
$$ 
and the shell-correction energy $E_{\rm shell}$ as a function of
the single-particle energies $\varepsilon_\nu$ by 
$$ 
E_{\rm shell} = \sum_\kappa \varepsilon_\kappa (n_\kappa - \bar n_\kappa)\,\, ,
\eqno(2.16)
$$
where the occupation probabilities $n_\kappa$ were defined in
(2.9), whereas the quantities $\bar n_\kappa$ represents smooth occupation 
probabilities in a cluster in which the shell structure is washed out using 
the Strutinsky prescription [13].

In our simple phenomenological model, we 
represent the distribution of the positive surplus charge
$[\rho_{\rm ion}(r) - \rho_e(r)]$ by a~simple ansatz: From
classical electrodynamics we know that the surplus charge ought
to be localized at the outer surface of the metallic bubble. We,
therefore, assume the total (positive) surplus charge $qe_0$ to
be distributed in a~thin layer of thickness $\varepsilon$ along
the outer surface 
$$ 
[\rho_{\rm ion}(r) - \rho_e(r)] = \delta\rho \cdot\theta_0
    (r - (R_1-\varepsilon)) \, \theta_0(R_1 - \varepsilon) \,\, ,
\eqno(2.17)
$$
where
$$
   \delta\rho = {3qe_0\over 4\pi[R^3_1 - (R_1 - \varepsilon)^3]}\,\, .
$$
Calculating the Coulomb energy (2.15) for this distribution we
obtain 
$$ 
  E_{\rm Cb} = {(4\pi\delta\rho)^2 R^5_1\over 3}
 \left\{{1\over 5}\left[1 - \left(1 - {\varepsilon\over
   R_1}\right)^5 \right] - {1\over 2} \left(1 - {\varepsilon\over
   R_1}\right)^3 \left(2{\varepsilon\over R_1} - {\varepsilon^2
     \over R^2_1}\right)\right\}\,\, .
\eqno(2.18)
 $$
Up to terms of the order of $\left({\varepsilon\over
R_1}\right)^2$, this can be written 
$$ 
E_{\rm Cb} = {q^2e^2_0\over 2R_1} \left[1 + {1\over 3} {\varepsilon\over R_1}
+ 0\left({\varepsilon^2\over R^2_r}\right)\right]\,\, .
 \eqno(2.19)
$$ 
The term of order $\left({\varepsilon\over R_1}\right)^0$ in
(2.19) represents the Coulomb energy in the limit that the
surplus charge is located in an infinitely thin layer at the
outer surface. One sees from (2.19) that the Coulomb energy
increases if the surplus charge is distributed homogeneously in
a~layer of finite thickness $\varepsilon \ll R_1$.
Consequently, we used the Coulomb energy in zeroth order of
${\varepsilon\over R_1}$ in our calculations.

The total energy of the spherical metallic bubble has thus the form
$$
    E_{\rm tot} = -\tau \cdot {4\pi\over 3} (R^3_1 - R^3_2) + \sigma 4\pi
    (R^2_1 + R^2_2) + {q^2e^2_0\over 2R_1} + E_{\rm shell}(f)\,\, .
\eqno(2.20)
$$
Subtracting from this expression the energy of a~compact
spherical cluster of the same charge $qe_0$ and the same volume
we obtain
$$
 \Delta E(f;q):~= \Delta E_{\rm LD}(f;q) + \Delta E_{\rm shell} (f;n-q)\,\, ,
\eqno(2.21)
$$
where
$$
\Delta E_{\rm LD}(f;q) = 4\pi\sigma \left(R^2_1 + R^2_2 - R^2_0\right)
 + {q^2e^2_0\over 2} \left({1\over R_1} - {1\over R_0}\right)
 \eqno(2.21')
 $$
 and
 $$
 \Delta E_{\rm shell}(f;n-1) = E_{\rm shell}(f;n-q) - E_{\rm shell}(0;n-q)
 \,\, .
 \eqno(2.21'')
 $$
We note that the single particle potential (2.12) becomes
a~simple central square-well in the limit $R_2 \rightarrow 0$
(i.e. $f\rightarrow 0$). Thus the shell correction energy
 $E_{\rm shell} (f = 0)$ is obtained by substituting the eigenvalues
 $\varepsilon_\kappa (f = 0)$ in a~simple central well with
 infinite wall at $r = R_0$. The occupation probabilities have to
 be chosen in each case according to Strutinsky's prescription.

The 1$^{\rm st}$ term and 2$^{\rm nd}$ term of (2.21') are simple
functions of the parameter $f$ due to the relations
$$
    R_1 = R_0 \cdot \left({1\over 1-f}\right)^{1/3} \,\, ,
\eqno (2.22)
$$
$$
     R_2 = R_0 \cdot \left({f\over 1-f}\right)^{1/3} \,\, .
\eqno(2.22')
$$
The radius $R_0$ of the compact spherical cluster is related to
the number of $n$ of atoms by
$$
  R_0 =\,\, \hskip -3pt\buildrel\circ\over r_s \, n^{1/3}
\eqno(2.23)
$$
with the radius parameter given in (2.2).

The surface tension $\sigma$ and the radius $\hskip
-3pt\buildrel\circ\over r_s$ of the Wigner-Seitz cell are thus
the only parameters which specify a~given metal in our model.

The difference $\Delta E_{\rm LD}(f;q)$ between the LD-energies of
the compact cluster and the bubble (Eqn. (2.21')) is seen to
consist of a~positive term describing the increase of the
surface energy and a~negative term which represents the
reduction of the repulsive Coulomb energy. As far as the
,,macroscopic'' part of the energy is concerned, a~preference
for the bubble geometry may only occur through the reduction of
the Coulomb energy which increases with the surplus charge $q$
of the metal cluster. For a~given metal and a~given ,,size'' $n$
of the cluster there will thus be a~critical value $q_c$ of the
charge at which the (spherical) bubble becomes the configuration
of lower energy. It will depend on the type of the metal (i.e.
on the value of the surface tension $\sigma$ and the radius
parameter $\hskip -3pt\buildrel\circ\over r_s$) whether the
cluster is still stable versus fission or emission of atoms at
this value of the charge.

The difference $\Delta E_{\rm shell}$ of the shell correction terms
can be positive or negative. If the number $(n-q)$ of valence electrons
happens to be a~magic number for the bubble geometry, and not
for the compound spherical form, $\Delta E_{\rm shell}$ will be
a~negative number and thus favour the formation of a~bubble
a.v.v. 

The question whether, for given charge, the bubble has a lower
energy than the compact sphere, depends sensitively on the
value of the surface constant $\sigma$ and the radius parameter 
$\buildrel\circ\over r_s$. More precisely, it depends on the
"fissility" parameter which is defined to be the ratio of the
Coulomb energy $E_{\rm Cb}$ and twice the surface energy $E_S$.
The factor 2 is inserted in order to retain the definition used
in nuclear physics:
$$
 X = {E_{\rm Cb}(f)\over 2E_S(f)} = {q^2e^2_0\over 
 16\pi\sigma\cdot R_1 (R^2_1+R^2_2)} = X_0 \cdot 
 \left({1 - f\over 1 + f^{2/3}}\right) \,\, ,
\eqno(2.24)
$$
with $X_0$, the fissility parameter of a compact sphere,
being defined by
$$
 X_0 = {E_{\rm Cb}(f=0)\over 2E_S(f=0)} = {q^2e^2_0 \over 
 16\pi\sigma \buildrel\circ\over r^3_s n}\,\, .
\eqno(2.25)
$$
The difference $\Delta E_{\rm LD}$ between the LD-energy of the
spherical bubble and of the compact spherical cluster (see
(2.21')) measured in units of $2E_S(f = 0)$ is given by the
function 
$$
F: = {\Delta E_{\rm LD}(f;q)\over 2 E_S(f=0)} = {1 \over 
 2(1-f)^{2/3}} \left[1 + f^{2/3} - (1 - f)^{2/3}\right] - X_0
 \left[1 - (1 - f)^{1/3}\right]\,.
\eqno(2.26)
$$
of the hole parameter $f$ ($0 \leq f < 1$), the 1$^{\rm st}$
term on the r.h.s. is positive and the 2$^{\rm nd}$ term is
negative. Clearly, for large enough $X_0$, the function $F$
becomes negative, i.e. the bubble shape corresponds to a~lower
energy. The stationarity condition
$$
 {\partial F\over \partial f} = 0
\eqno(2.27)
$$
has the explicit form
$$
 1 + f^{1/3} - X_0 \, f^{1/3}(1 - f) = 0\,\, .
\eqno(2.28)
$$
Solving Eq. (2.28) one can find that the metallic cluster with charge
$q\cdot e_0$ and the atom number $n$ have their lowest energy for a spherical
bubble shape if the fissility parameter $X_0 < X_0^{cr}$= 3,4 .

A crucial question is the dependence of the total energy on
deformation: It can be easily shown that the LD-energy of
a~bubble decreases as a~function of deformation because the
decrease of the repulsive electrostatic energy turns out to be
greater than the increase of the surface energy. A~spherical
bubble may, however, be stabilized against deformation by shell
effects. If the $(n-q)$ valence electrons of the charged metal
bubble correspond to a~closed shell configuration in the
spherically symmetric potential (2.12), the shell energy yields
an additional binding of a~couple of eV. As one deforms the
bubble the absolute value of this negative shell energy
decreases as a~function of deformation. In this way a~barrier
against fission is produced. The calculation of the total energy
of the bubble as a~function of the deformation implies that we 
determine the eigenvalues in a~deformed bubble potential. This
is a~difficult task if we were to tackle it in full generality.
There is, however, a~family of deformed shapes which can be
transformed into a~spherical shape by a~scaling transformation.
For this "scaling model", it is relatively simple to calculate
the eigenvalues:

Assume that the outer ($S_1$) and inner ($S_2$) surface of the
deformed bubble are concentric spheroids with the half-axes
$a_{1(2)}$ and $c_{1(2)}$
$$
 S_{1(2)}:\,\,\, {x^2 + y^2\over a^2_{1(2)}} + {z^2\over c^2_{1(2)}} = 1
\eqno(2.29)
$$
enclosing the constant volume ${4\pi\over 3}(R^3_1 - R^3_2)$
$$
 R^3_0 = R^3_1 - R^3_2 = a^2_1\, c_1 - a^2_2c_2\,.
\eqno(2.30)
$$
An infinite square well with the boundaries (2.29)
$$ 
\hat V(x,y,z) = - V_0\, \eta_0(\vec x)\,\, ,
\eqno(2.31)
$$
where
\begin{eqnarray*}
&& \eta_0(\vec x) = 
 \left\{\begin{array}{ll}
1 & {\rm for}~ \vec x~{\rm \in volume}~ \Omega~{\rm
    enclosed~by}~S_1~{\rm and}~ S_2 \\ 
  -\infty & {\rm otherwise}
\end{array}\right.
\end{eqnarray*}
transforms into a simple spherically symmetry potential of the
type (2.12) which depends on the scaled variable
$$
 \xi = \lambda x;~~~~~~~~\eta = \lambda y;~~~~~~~~~\zeta = \mu z
\eqno(2.32)
$$
or the corresponding polar variables $\rho,\vartheta,\varphi$
$$
\xi = \rho \, \sin\vartheta \,\cos\varphi;~~~~~~~~ 
\eta = \rho \, \sin\vartheta\, \sin\varphi;~~~~~~~
\xi = \rho\cos\vartheta\,.
\eqno(2.32')
$$
The 2 scaling parameters $\lambda$, $\mu$ are related to each
other by the constraint (2.25) which takes the form
$$
 \lambda^2 \mu = 1\,\, .
\eqno(2.33)
$$
Thus they can be expressed by a single deformation parameter.
Using the deformation parameter $\delta$ introduced
by S. G. Nilsson in Ref. [25] the scaling parameters are given as 
a function of $\delta$ by
$$
\lambda = \left ( {1 + \frac{2}{3}\delta \over 1 - \frac{4}{3}\delta }
          \right )^\frac{1}{6} \,\,,\,\,\,\,
\mu = \left ( {1 - \frac{4}{3}\delta \over 1 + \frac{2}{3}\delta }
          \right )^\frac{1}{3} \,\, .
\eqno(2.34)
$$
In the scaled variables, the Schr\"odinger
equation takes the form
$$
(\hat H_0 + \hat H_1) \varphi_\nu = \varepsilon_\nu\,\varphi_\nu \,\, .
\eqno(2.35)
$$
The hamiltonian $\hat H_0$ is spherically symmetric in the $\xi,\eta,\zeta$
coordinates
$$ 
 \hat H_0 = -{\hbar^2\over 2M_\delta} \, \Delta_{\xi\eta\zeta} + V(\rho)\,\, ,
\eqno(2.36)
$$
where $\Delta_{\xi\eta\zeta}$ is the Laplacian operator in $\xi,\eta,\zeta$
space and
$$
M_\delta = M (1 + \frac{2}{3}\delta)^\frac{2}{3}
             (1 - \frac{4}{3}\delta)^\frac{1}{3} \,\, .
\eqno(2.37)
$$
The scaled square well potential is now spherical
$$
 V(\rho) =  \left\{ 
 \begin{array}{ll}
 -V_0     & {\rm for}~~R_2 < \rho < R_1 \\
 + \infty & {\rm otherwise}
 \end{array}\right. \,\,.
\eqno(2.38)
$$
The term $\hat H_1$ in (2.31) represents the deformation dependent part
of the hamiltonian
$$
 \hat H_1 = \frac{2}{3}\delta {\hbar^2\over 2M} \, \left (
   2 {\partial^2\over \partial \zeta^2} - {\partial^2\over \partial \xi^2}
   - {\partial^2\over \partial \eta^2} \right ) \,\, .
\eqno(2.39)
$$

For determining of the stability of spherical bubble with respect
to elongation, we only need to consider small
deformations $\delta$. Consequently, we may treat $\hat H_1$ as
a~perturbation. This means that the single-particle energy
$\varepsilon_\nu$ can be approximately obtained as a~function of
the eigenenergies $\buildrel\circ\over\varepsilon_\nu$ and
eigenfunctions $\buildrel\circ\over\varphi_\nu$ of the
Hamiltonian $\hat H_0$ as
$$
 \varepsilon_\nu \approx \buildrel\circ\over \varepsilon_\nu +
 \langle \buildrel\circ\over\varphi_\nu|\hat H_1|
  \buildrel\circ\over \varphi_\nu\rangle\,.
\eqno(2.40)
$$
For larger deformations, one has to diagonalize $\hat H_1$
in the basis of the s.p. states $\buildrel\circ\over\varphi_\nu$.
In all the results shown in Section 3, the eigenvalues
$\varepsilon_\nu$ were determined by diagonalization of
$\hat H_1$ in a~sufficiently large subspace of s.p. states
$\buildrel\circ\over \varphi_\nu$.

The matrix-elements of $\hat H_1$ can be easy evaluated
using the following artifice
$$
  2 {\partial^2\over \partial \zeta^2} - {\partial^2\over \partial \xi^2}
  - {\partial^2\over \partial \eta^2} = \frac{1}{8} \left [
    \Delta_{\xi\eta\zeta} , \left [ \Delta_{\xi\eta\zeta}, 
    2\rho^2 P_2(cos\theta) \right ] \right ] =
    \frac{1}{8} \left (\frac{2 M_\delta}{\hbar^2}\right )^2
    \left [\hat H_0 , \left [\hat H_0 , 2 \rho^2 P_2(cos\theta)\right ] 
    \right ] \,\, .
\eqno(2.41)
$$
where $P_2$ is the Lagrange polynomial. Simple algebra leads to the expression
$$
 \langle \buildrel\circ\over\varphi_\nu|\hat H_1|
 \buildrel\circ\over \varphi_\mu\rangle = \frac{\delta}{6} 
 \left (\frac{2 M_\delta}{\hbar^2} \right ) (\buildrel\circ\over e_\nu -
 \buildrel\circ\over e_\mu)^2 \langle \buildrel\circ\over\varphi_\nu|
 \rho^2 P_2(cos\theta) |\buildrel\circ\over \varphi_\mu\rangle \,\, ,
\eqno(2.42)
$$
The matrix elements of $\rho^2$ where evaluated numerically, while 
the matrix elements of spherical harmonics are expressed in terms
of Clebsch-Gordan coefficients.

\section{Results}

The calculation was performed for metallic agglomerates of sodium ($Na$) 
and of antimonium ($Sb$). The liquid drop parameters are the same as in
Ref. [21] where the fission of charged clusters was discussed.
For $Na$ clusters we have used the set
$$
\buildrel\circ\over r_s = {\rm 2.070\, \AA} \,\,,\,\, 
     \tau = {\rm 0.03017\, eV} \,\,,\,\, 
     \sigma = {\rm 0.01894\, eV}
$$
and for $Sb$ clusters
$$
\buildrel\circ\over r_s = {\rm 1.130\, \AA} \,\,,\,\,
     \tau = {\rm 0.4552\, eV} \,\,,\,\, 
     \sigma = {\rm 0.02474\, eV} \,\, .
$$

The results for the charged sodium clusters are presented in Fig. 1
The number $n$ of atoms in clusters vary from 700 to 10000, while
the charge number $q$ changes from 30 to 200. The liquid drop estimate 
of the binding energy ($E_{LD}$, Eq. 2.1), the energy gain ($\Delta E_{LD}$, 
Eq. 2.21') with respect to the energy of a compact spherical cluster, 
the fissility parameter ($X$, Eq. 2.24) and the equilibrium hole fraction 
($f$, Eq. 2.4) are drawn. It is seen in the Fig. 1 that only massive
and relatively highly charged $Na$ clusters favour the bubble solution.
The average energy gain is around 0.3 eV per atom. The hole fraction
parameter $f$ varies from 0.3 to almost 0.9 and the hole in the bubble is
larger the higher is the charge of the cluster. Unfortunatelly the
fissility parameter $X$ is always larger than one. This means that
there is no fission barrier in the liqud model and there is a little
hope that the shell effect is large enough to stabilize the sodium bubble.

A more optimistic situation exists for the $Sb$ clusters, which have
a smaller Wigner-Seitz radius constant. In Fig. 2 we plotted analogous
results as in Fig. 1. One can see that already
small and not highly charged clusters prefer the bubble configuration.
The energy gain with respect to the compact shape could even reach 
1 eV per atom, also the binding energies per atom are few times larger
than for Na bubbles. But the fissility parameter $X$ is here also
larger than one. 
One can conclude from the LD results presented in Figs 1 and 2 that
within the LD model
stable bubble clusters are rather improbable. Only the shell effects
may stabilize them with regard to fission. 

The electronic
scheme obtained for the infinite spherical square well potential
(2.12) is plotted in Fig. 3 as a function of the hole fraction $f$ (2.4).
The energy unit used takes into account that the eigenvalues of
the infinite square well scale with $n^{-2/{3}}$. The radius constant
for $Sb$ is used here.
It is seen in the figure that for larger $f$ the orbitals with 
the node number $n > 1$ corresponds to much higher energies than 
those with $n=1$ and the levels are well separated from the others.
This effect leads to strong shell effects. A similar tendency was also
observed in the shifted harmonic oscillator [23]. New magic numbers
corresponding to the orbitals with $l$=0, 1, 2, 3, .... are found:
$$
{\cal M}_{bubb.} = 2, 8, 18, 32, 50, 72, 98, 128, 162, 200, 242, 288, 338,
             392, 450, 512, 578, 648 ....
$$
These magic numbers are quite different from those observed in the
compact spherical clusters [26] :
$$
{\cal M}_{comp.} = 2, 8, 20, 34, 58, 92, 138, 168, 254, 338, 438, 440, 
542, 556, 676 .....
$$
The Strutinsky shell correction energy for the $Sb$ bubble cluster with the
hole fraction $f$=0.7 is plotted as a function of $n$ in Fig. 4.
The shell correction energy for a magic number of valence electrons in
the bubble cluster is negative
and reaches even -3 eV for n=648. The magnitude of the shell correction
is comparable with that for supershells for heavy clusters with
1074 $\leq n \leq$ 3028 discussed by Brack in Ref. [26]. Such a huge
shell effect for magic clusters should protect them against fission. 
It can be seen in Fig. 5, where the fission barrier for the magic
bubble cluster $_{210}Sb^{+10}$ is analysed. In the upper l.h.s. part
of Fig. 5, the liquid drop energy is drawn as a function of the hole 
fraction $f$. A pronounced bubble minimum is observed at $f \approx$
0.4 . The liquid drop part of the energy at $f$=0.4 is plotted as a function
of deformation parameter $\delta$ in the upper r.h.s. part of Fig. 5.
As one can expect there is no fission barrier in this case.  
The two lower graphs represent the shell energy (2.16) and the fission 
barrier beeing the sum of the LD part and the shell correction (2.13) 
as a function of $\delta$.
It is seen that $E_{shell}$=-2,7 eV for spherical configuration and its
magnitude decreases with deformation and oscillates. This effect produces
a prononced fission barrier which makes fission of $_{210}Sb^{+10}$
rather improbable.

\section{Summary and discussion}

We have investigated the shell correction energy of charged spherical 
bubble clusters. The lower limit of mass and charge numbers, where
bubble clusters begin to exist depends sensitively on the choice of 
the LD parameters.
The investigation shows that promising candidates for bubble
structure are $Sb$ clusters with a magic number of itinerant 
electrons in the charged agglomerate. The charge number is found to be
a fraction of 0.04 to 0.15 of the
atom number. The $Na$ cluster should contain large atom number,
preferably a few thousands and they should be highly charged 
($q \approx$ 100).

We have found strong shell effects which may give rise to
shell energies of up to -4~eV for certain magic numbers.  By
calculating the LD-energy for deformed bubbles and the deformation
dependence of the shell effect, we found that the fission
barriers are of the same order of magnitude as the
shell-correction energy for the spherical bubble solution. In
favorable cases of magic numbers, this is sufficient to reduce
significantly the probability for spontaneous fission.
The origin of the shell effects is the high degeneracy of
orbitals with large angular momentum and the rapid energetic
increase of radial oscillation modes as a~function of increasing
bubble radius.  

We expect that similarly big shell effects could exist in the
clusters in which the outer layer from other material is added.
If would exist the potential barrier between the inner and outer 
material then the electrons in the outer metallic layer could feel the magic 
numbers like those found for the bubble cluster. It would be a great 
challenge for experimentalists to produce (by epitaxy?) such objects.

\bigskip\bigskip\noindent
{\bf Acknowledgements}

\bigskip\noindent

Krzysztof Pomorski gratefully acknowledges the warm hospitality extended
to him by the Theoretical Physics Group of the Technical University in
M\"unchen as well as to the Deutsche Forschung Gemeinshaft for granting 
a~guest professor position which enabled us to perform this research.

\newpage       
\noindent
{\bf References}

\begin{enumerate}

\item Knight, W. D., Clemenger, K., de Heer, W. A., Saunders, W.
A., Chou, M. Y., Cohen, M. L.: Phys. Rev. Lett. {\bf 52}, 2141
(1984) 

\item Knight, W. D., de Heer, W. A., Clemenger, K., Saunders, W.
A.: Solid State Commun. {\bf 53}, 44 (1985)

\item Ekardt, W.: Phys. Rev. {\bf B29}, 1558 (1984)

\item Beck, D. E.: Solid State Commun. {\bf 49}, 381 (1984)

\item Chou, M. Y., Cleland, A., Cohen, M. L.: Solid State
Commun. {\bf 52}, 645 (1984)

\item Brack, M., Genzken, O., Hansen, K.: Z. Phys. {\bf D21}, 65
(1991); ibid. {\bf 19}, 51 (1991)

\item Nishioka, H., Hansen, K., Mottelson, B. R.: Phys. Rev.
{\bf B42}, 9377 (1990)

\item Koch, E.: Phys. Rev. Lett. {\bf 76} (1996) 2678

\item Clemenger, K.: Phys. Rev. {\bf B32}, 1359 (1985)

\item Reimann, S. M., Brack, M., Hansen, K.: Z. Phys. {\bf D28},
      235 (1993)

\item Frauendorf, S., Pashkevich, V. V.: Z. Phys. {\bf D26}, 98
      (1993) 

\item Hamamoto, I., Mottelson, B. R., Xie, H., Zhang, X. Z.: Z.
      Phys. {\bf D21}, 163 (1991)

\item Strutinsky, V. M.: Sov. J. Nucl. Phys. {\bf 3}, 449 (1967);
      Nucl. Phys. {\bf A95}, 420 (1967); ibid. {\bf A122}, 1 (1968) 

\item Ekardt, W., Penzar, Z.: Phys. Rev. {\bf B38}, 4273 (1988)

\item Lauritsch, G., Reinhard, P.-G., Meyer, J., Brack, M.:
      Phys. Lett. {\bf A160}, 179 (1991)

\item Bj\"ornholm, S., Borggreen, J., Echt, O., Hansen, K.,
      Pedersen, J., Rasmussen, H. D.: Phys. Rev. Lett. {\bf 65}, 1627
      (1990); Z. Phys. {\bf D19}, 47 (1991) with ref. to earlier work

\item Br\'echignac, C., Cahuzac, Ph., Carlier, F., Trutos, M.
      de, Roux, J. Ph.: Phys. Rev. {\bf B47}, 2271 (1993), with ref.
      to earlier work

\item Br\'echignac, C., Busch, H., Cahuzac, Ph., Leygnier, J.:
      J. Chem. Phys. {\bf 101}, 6992 (1994)

\item Br\'echignac, C., Cahuzac, Ph., Carlier, F., Trutos, M.
      de, Roux, Leygnier, J. Ph.: J. Chem. Phys. {\bf 102(2)}, 763
      (1995) 

\item Garcias, F., Lombard, R. J., Barranco, M., Alonso, J. A.,
      L\'opez, J. M.: Z. Phys. {\bf D33}, 301 (1995)

\item Gross, D.H.E., Hervieux, P.A.: Z. Phys. {\bf D35}, 27 (1995), \\
      Hervieux, P.A., Gross, D.H.E.: Z. Phys. {\bf D33}, 295 (1995), \\
      Gross, D.H.E., Madjet, M.E., Schapiro, O.: Z. Phys. {\bf D39},
      75 (1997)

\item  Myers, W.D., Swiatecki, W.J.: Ark. Fys. {\bf 36}, 343 (1966),\\
       Myers, W.D., Swiatecki, W.J.: Nucl. Phys. {\bf A601}, 141 (1996)

\item Dietrich, K., Pomorski, K.: accepted for publication in
      Phys. Rev. Lett. and Nucl. Phys. A

\item Gunnarsson, O., Lundquist, B. I.: Phys. Rev. {\bf B13},
      4274 (1976)

\item Nilsson, S. G.: Mat. Fys. Medd. Dan. Vid. Selsk. {\bf 29},
      no. 16 (1955)

\item Brack, M.: Rev. Mod. Phys. {\bf 65}, 677 (1993)

\end{enumerate}

\newpage
\noindent
{\bf Figures captions:}

\begin{enumerate}
\item Liquid drop estimate of the binding energy ($E_{LD}$, Eq. 2.1),
      energy gain ($\Delta E_{LD}$, Eq. 2.21') with respect to 
      the energy of a compact spherical cluster, fissility parameter
      ($X$, Eq. 2.24) and the equilibrium hole fraction ($f$, Eq. 2.4) 
      for the charged ($q\cdot e_0$) sodium cluster with the bubble structure
      as function of $n$.
\item The same as in Fig. 1 but for  charged clusters of antimonium.
\item Electronic level scheme as a function of the hole fraction 
      $f= (R_2/{R_1})^3$ for the infinite spherical square well (2.12).
      The energy unit used takes into account that the eigenvalues of
      the infinite square well scale with $n^{-2/{3}}$. The radius constant
      for $Sb$ cluster is used here.
\item Shell correction energy for the $Sb$ bubble cluster with the
      hole fraction $f$=0.7 as a function of $n$.
\item Liquid drop energy of $_{210}Sb^{+10}$ as a function of the hole
      fraction $f$ (upper l.h.s. figure) and the deformation parameter
      $\delta$ (upper r.h.s). The two lower graphs represent the shell
      energy (2.16) and the fission barrier (2.13) as a function of $\delta$.
\end{enumerate}

\newpage
\begin{figure}
\epsfysize=180mm \epsfbox{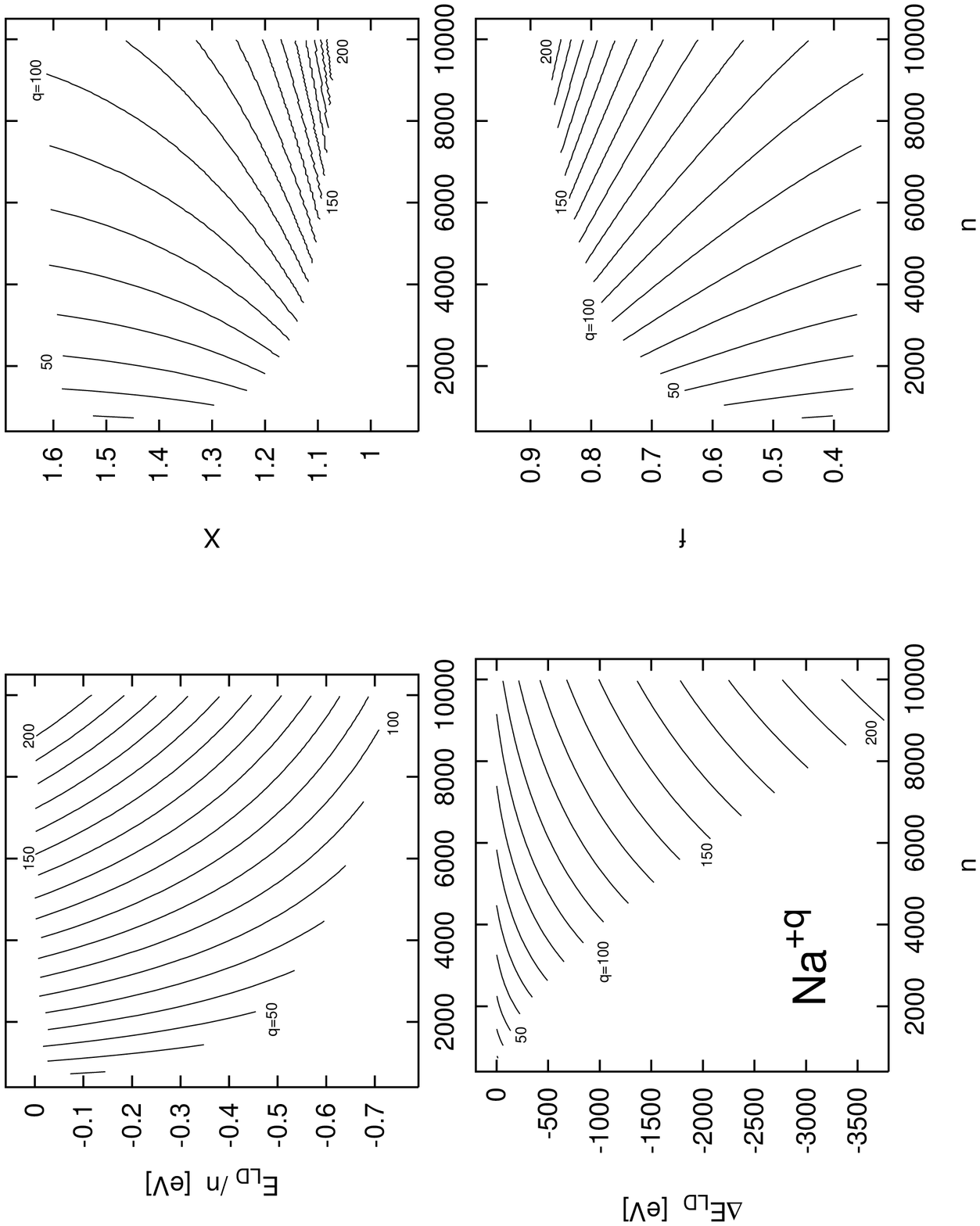}
\caption{ }
\end{figure}

\newpage
\begin{figure}
\epsfysize=180mm \epsfbox{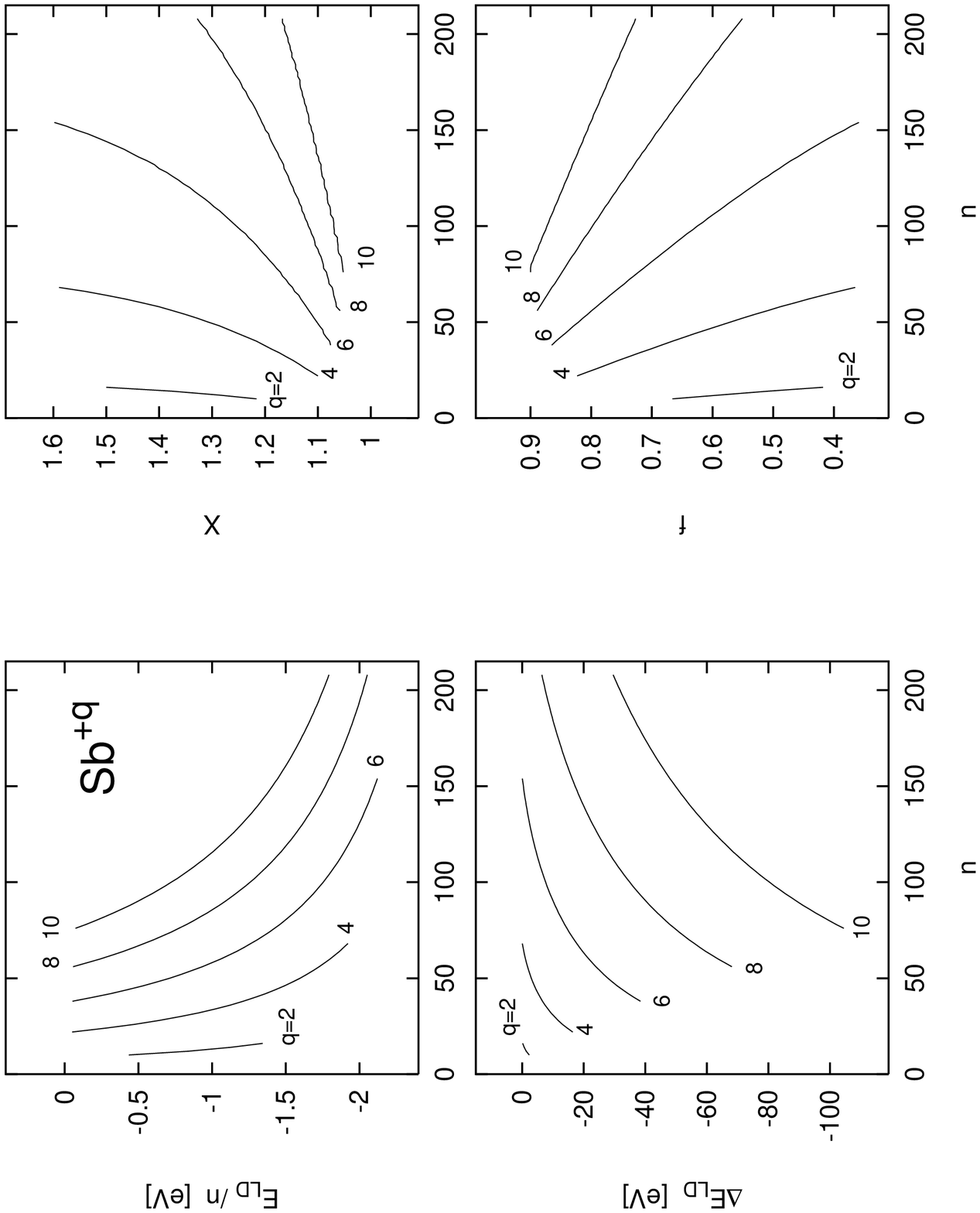}
\caption{ }
\end{figure}

\newpage
\begin{figure}
\epsfysize=210mm \epsfbox{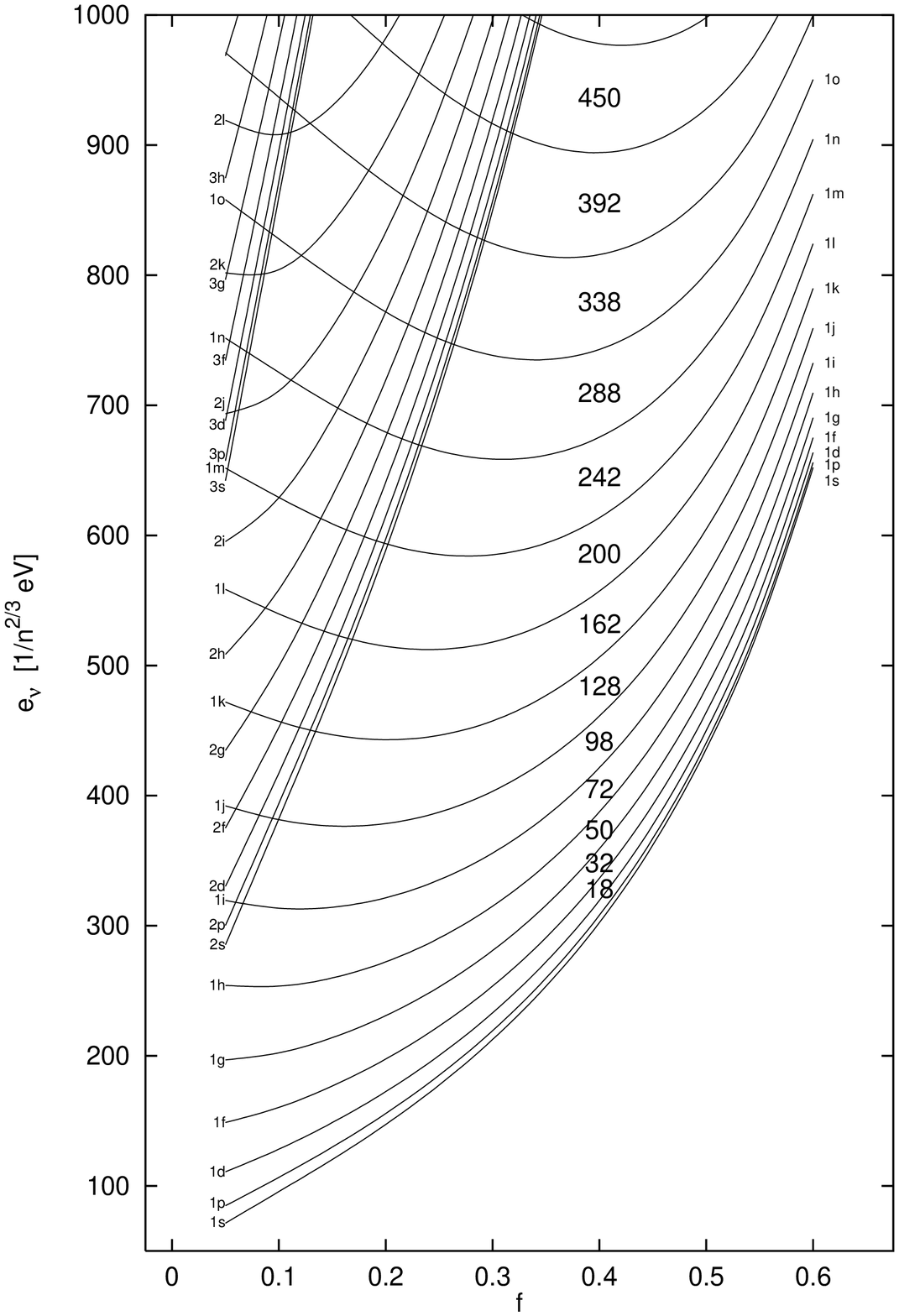}
\caption{ }
\end{figure}

\newpage
\begin{figure}
\epsfysize=180mm \epsfbox{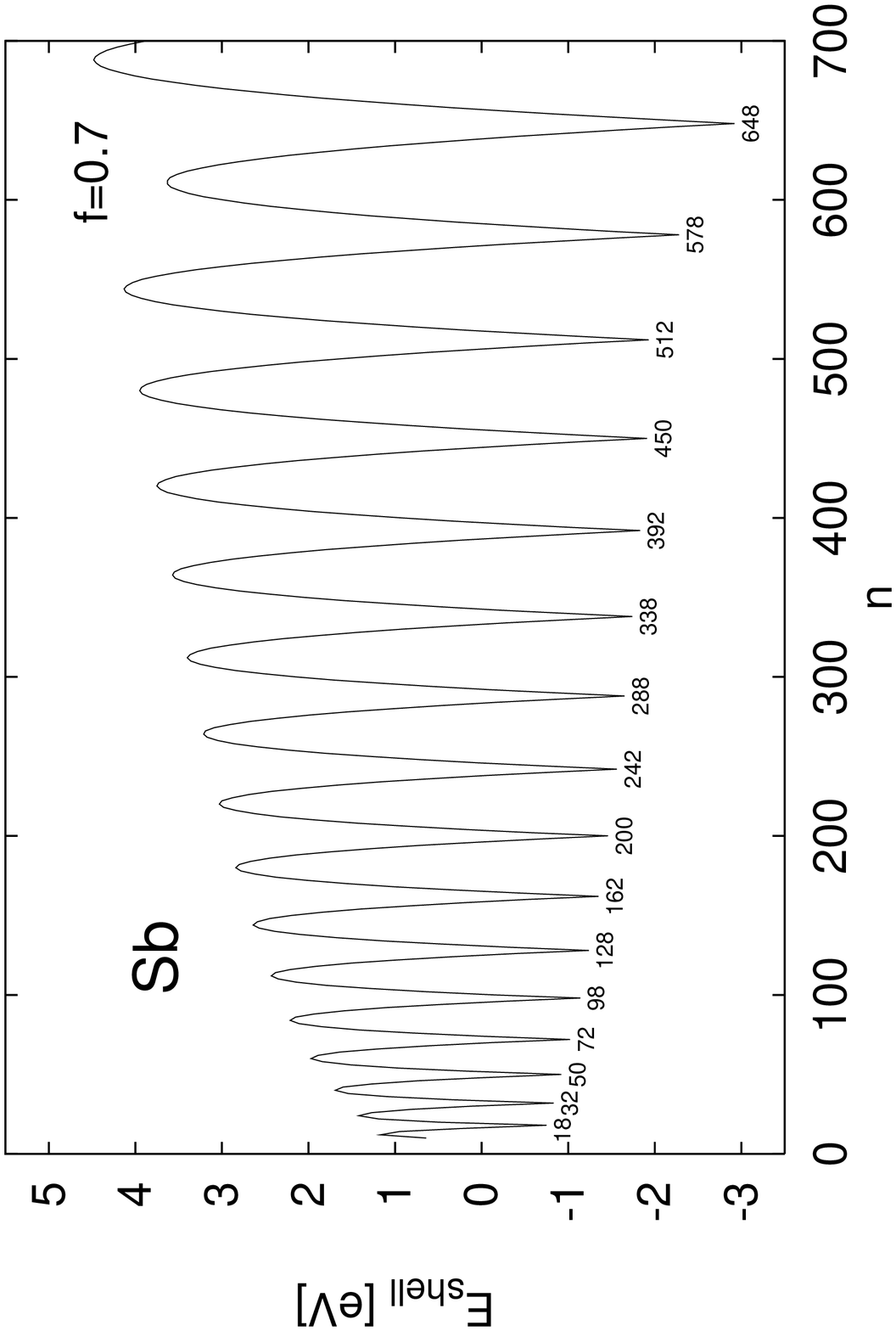}
\caption{ }
\end{figure}

\newpage
\begin{figure}
\epsfysize=180mm \epsfbox{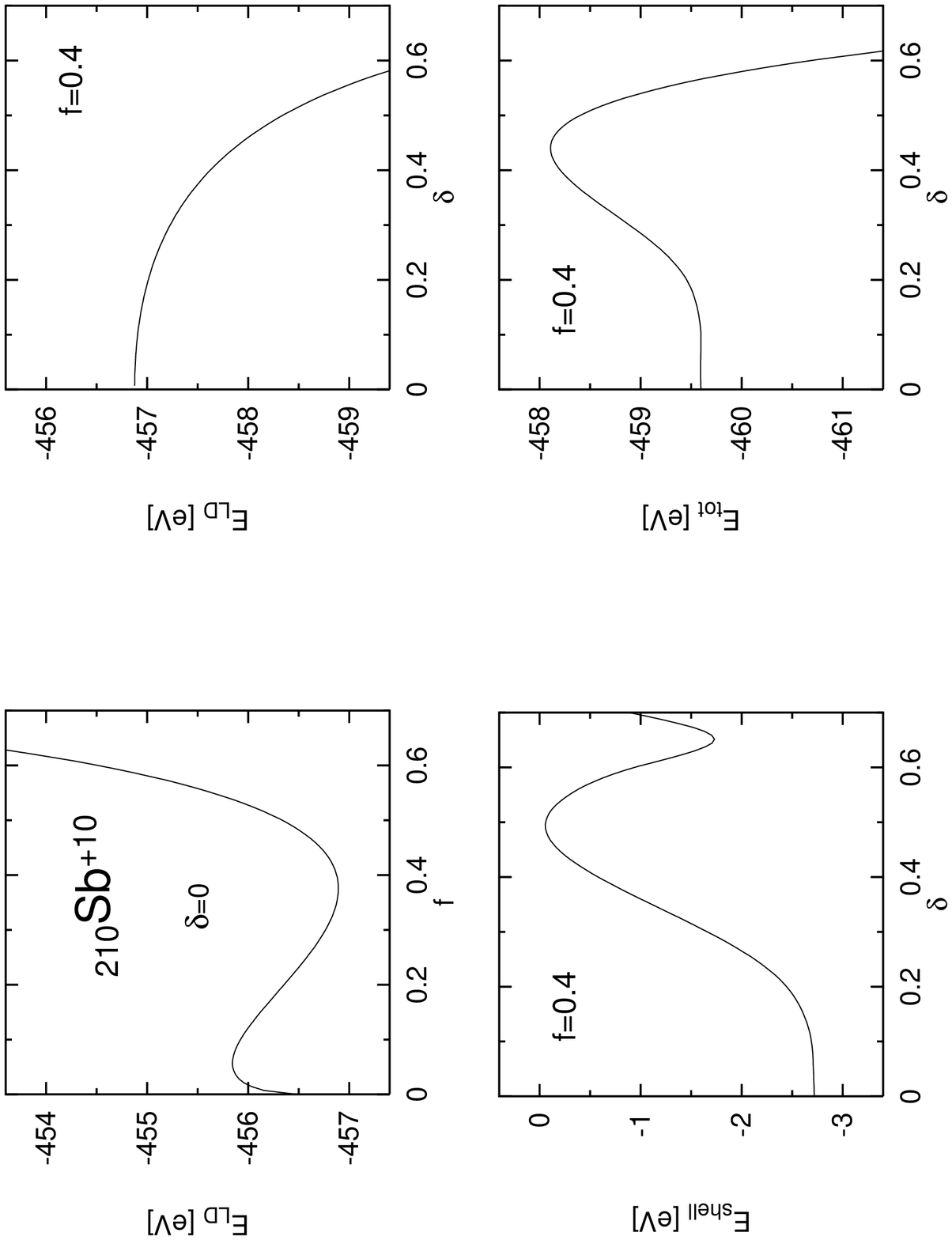}
\caption{ }
\end{figure}
\newpage
\end{document}